\documentclass[preprint,showpacs,preprintnumbers,aps,prd,superscriptaddress]{revtex4-2}

\usepackage{graphicx,color}
\usepackage{hyperref}
\usepackage{amsmath,amssymb}
\usepackage{url}
\usepackage{bbold}
\usepackage[normal]{subfigure}
\usepackage{array}

\usepackage{ORCIDinREVTeX}

\newcommand{\be}{\begin{equation}}
\newcommand{\ee}{\end{equation}}
\newcommand{\bea}{\begin{eqnarray}}
\newcommand{\eea}{\end{eqnarray}}

\newcommand{\bc}{\begin{center}}
\newcommand{\ec}{\end{center}}

\newcommand {\ba}{\begin{array}}
\newcommand {\ea}{\end{array}}
\newcommand{\ben}{\begin{enumerate}}
\newcommand{\een}{\end{enumerate}}

\usepackage{epsfig,graphicx}
\usepackage{bm}
\usepackage{dcolumn}
\allowdisplaybreaks

\newcommand{\avg}[1]{\left< #1 \right>} 

\begin{document}

\preprint{}

\title{On the triplet of holographic phase transition}

\author{Tran Huu Phat}
\email{thphat@live.com}
\affiliation{Vietnam Atomic Energy Commission, 
    59 Ly Thuong Kiet street, Hoan Kiem, Hanoi 100000, Vietnam.}
\author{Toan T. Nguyen}
\orcid{0000-0002-6331-2453}
\email{toannt@hus.edu.vn, toannt@vnu.edu.vn}
\affiliation{Key Laboratory for Multiscale Simulation of Complex Systems, and Department of Theoretical Physics, \\
University of Science, Vietnam National University $-$ Hanoi,
334 Nguyen Trai street, Thanh Xuan, Hanoi 100000, Vietnam.}

\begin{abstract}
We start from an Einstein $-$ Maxwell system coupled with a charged scalar field in Anti$-$de Sitter space$-$time.
In the setup where the pressure $P$ is identified with the cosmological constant,
the AdS black hole (BH) undergoes the phase transition from small to large BHs,
which is similar to the transition from liquid to gas in the van der Waals theory.
Based on this framework, we study the triplet of holographic superconducting states, consisting of ground state and two lowest excited states.
Our numerical calculations show that the pressure variation in the bulk creates a mechanism in the boundary which causes changes in the physical properties of excited states, namely:
a) when the pressure $ P $ is higher than the critical pressure ${P_c}$ ($ P > {P_c} $) of the phase transition from small to large BHs the ground state and the first excited state are superconducting states while the second excited state is the normal one.
However, at lower pressure, $P \le P_c$, the ground state is solely the superconducting state. We conjecture that the precedent phenomena 
could take place when  the scalar field in the bulk is replaced by other matter fields.
\end{abstract}

 
\maketitle

\section{Introduction}\label{intro}
More than twenty years ago the AdS/CFT duality \cite{maldacena1999large} and its related GKPW relation \cite{witten1998anti, gubser1998gauge} have provided a new theoretical framework  
for finding out various holographic superconductors \cite{zaanen2015holographic, ammon2015gauge} which were stimulated by a series of papers \cite{gubser2008breaking, hartnoll2008building, hartnoll2008holographic}:  
the holographic superconductor was built up by means of a simple Einstein – Maxwell theory coupled to a charged scalar field which yielded a holographically dual description of superconductor. 
The authors indicated that the scalar condensate could be identified to high temperature superconductor. 
It is worth to emphasize that most of the obtained superconductors have been studied in the probe approximation. 
Then the gravity theory depends on three parameters: the charge $ Q $, the mass $ M $ and the cosmological constant $ \Lambda $. 
The character of superconductors at different values of charge $ Q $ was considered in Ref. \cite{horowitz2009zero} 
while the main goal of Ref. \cite{horowitz2008holographic} is to explore how the properties of superconductors depend on the scalar mass of the BH and, 
as a consequence, several new properties were discovered.

All foregoing holographic superconductors are in the ground state. 
In recent years there exhibits a new trend studying holographic superconductors, in which the ground state arises together with other excited states \cite{li2020nonequilibrium}, 
namely, the number of excited states depends on the value of chemical potential $\mu $. 
For example, for $\mu  = 4.064$ there is only the ground states while $\mu  = 35$ there appear six excited states. 
Inspired by  this work, the main aim of the present paper is to explore the impacts of the topological charge on the holographic phase transitions of not only ground state, but also all related excised states and their conductivity.
To this end, let us proceed to the model of an Abelian Higgs field coupled to  a Maxwell field in the four – dimensional space$-$time Einstein gravity. 
The bulk action takes the form 
\begin{gather}
S = \frac{1}{{16\pi {G_N}}}\int {{d^4}x} \sqrt { - g} \left[ 
    {R - \frac{6}{{{L^2}}} - \frac{1}{4}F_{\mu \nu }^2 - {{\left| {{\partial _\mu } - iq{A_\mu }\psi } \right|}^2} - m{{\left| \psi  \right|}^2}}
    \right] ,
\end{gather}       
where ${G_N}$ is the Newton constant. 
In the non-condensed phase, solutions to the equation 
\begin{equation}\label{eq1}
\delta S = 0,
\end{equation}
are the Reissner – Nordstrom black hole (BH)
\begin{equation}\label{eq2}
d{s^2} =  - f\left( r \right)d{t^2} + \frac{{d{r^2}}}{{f\left( r \right)}} + {r^2}d\Omega _{2,k}^2,
\end{equation}
in which
\begin{equation}\label{eq3}
f\left( r \right) = k - \frac{{2M}}{r} + \frac{{{Q^2}}}{{{r^2}}} + \frac{{{r^2}}}{{{L^2}}},
\end{equation}
and
\begin{gather}
\begin{array}{l}
\psi \left( r \right) = 0,\\
\phi \left( r \right) = \displaystyle Q\left( {\frac{1}{r} - \frac{1}{{{r_0}}}} \right) .
\end{array}
\end{gather}
The horizon radius ${r_0}$ is the largest root of the equation
\begin{equation}\label{eq4}
f\left( {{r_0}} \right) = k - \frac{{2M}}{{{r_0}}} + \frac{{{Q^2}}}{{r_0^2}} + \frac{{r_0^2}}{{{L^2}}} = 0.
\end{equation}
In Eq. (\ref{eq2}), $d\Omega _{2,k}^2$ is the metric of a two sphere of radius $ 1 $. Note that the parameters $ M $ and $ Q $ are different from the mass and charge of BH by corresponding factors.\\
The Hawking temperature $ T $ and entropy $S$ of BH are, respectively:
\begin{subequations}\label{eq5}
	\begin{align}
	T = \frac{{f'\left( {{r_0}} \right)}}{{4\pi }}, \\
	S = \pi r_0^2.
	\end{align}
\end{subequations}
A breakthrough in this direction \cite{kubizvnak2012p, kubizvnak2017black} is to identify pressure with the cosmological constant  of BH
\begin{equation}
P =  - \frac{\Lambda }{{8\pi }} = \frac{3}{{8\pi {L^2}}}\,
\end{equation}
which leads to the analogy between the small – large BHs phase transition and the liquid – gas phase transition in the van der Waals theory for $ k>0 $. 
Then $ M $ becomes the enthalpy of the system and $ k $ is related to the topological charge 
\cite{tian2014holographic, tian2019topological}.
Consequently, the first  thermodynamic law of BH is extended \cite{lan2018phase}:
\begin{equation}\label{eq7}
dM = TdS + \omega d\varepsilon  + \phi dQ + VdP,
\end{equation}
where $\varepsilon  = 4\pi k$ is the topological charge and its conjugate potential is $\omega  = {r_0}/8\pi $, 
the charge $ Q $ conjugates to its potential $\phi  = Q/{r_0}$ and volume $V = \frac{4}{3}\pi r_0^3$ is conjugates to the pressure.

From Eqs. (\ref{eq3}), (\ref{eq4}) and (\ref{eq5}), one gets the isobaric specific heat
\begin{equation}
{C_{PQ\varepsilon }} = T{\left( {\frac{{\partial S}}{{\partial T}}} \right)_{P,Q,\varepsilon }} = \frac{{2\pi r_0^2\left( {32{\pi ^2}\Pr _0^4 + \varepsilon r_0^2 - 4\pi {Q^2}} \right)}}{{32\pi \Pr _0^4 - \varepsilon r_0^2 + 12\pi {Q^2}}},
\end{equation}
which yields critical values of various thermodynamic quantities for the transition from small to large BHs:
\begin{equation}\label{eq8}
{P_c} = \frac{{{\varepsilon ^2}}}{{1536{\pi ^3}{Q^2}}}~,~
{T_c} = \frac{{\sqrt 6 {\varepsilon ^{3/2}}}}{{96{\pi ^{3/2}}Q}}~,~
{r_c} = \frac{{2\sqrt {6\pi } Q}}{{\sqrt \varepsilon  }}.
\end{equation}
Expanding the results of Ref. \cite{phat2021holographic},
in this paper we consider the triplet of holographic phase transitions which associate with ground state, first excited state and second excited state.
Our main goal is to look for those effects which could occur for the triplet of holographic transitions in the process of the transition from  small to large BHs.
For simplicity and without losing generality, 
we set $ k = Q = q = 1 $ from now on.

The structure of this paper is as following. 
In Section \ref{2} we will briefly present all  basic materials of the AdS/CFT duality which will be employed in this paper. 
Section \ref{3} is devoted to the numerical computations associated with  the triplet of holographic superconductors and their physical properties.
The conclusion is presented in Section \ref{4}.

\section{Preparation }\label{2}

\subsection{Basic set up}

The first part of the basic set up is to build a framework  for numerically  calculating the triplet  of holographic phase transitions. First, as usual, we adopt the ansatz
\begin{equation}\label{eq9}
{A_\mu } = \left( {\phi \left( r \right),0,0,0} \right),\psi  = \psi \left( r \right).
\end{equation}
The mass of scalar field $ m $ is chosen above the Breitenlohner – Freedman bound \cite{breitenlohner1982stability}
\begin{equation}\label{eq10}
{m^2} =  - \frac{2}{{{L^2}}}.
\end{equation}
Inserting Eq. (\ref{eq9}) into Eq. (\ref{eq1}), one obtains
the equations of motion for matter field after some short derivations:
\begin{equation}\label{eq11}
\phi ''\left( r \right) + \frac{2}{r}\phi '\left( r \right) - \frac{{2{\psi ^2}\left( r \right)}}{{f\left( r \right)}}\phi \left( r \right) = 0,
\end{equation}
\begin{equation}\label{eq12}
\psi ''\left( r \right) + \left( {\frac{2}{r} + \frac{{f'\left( r \right)}}{{f\left( r \right)}}} \right)\psi '\left( r \right) + \left( {\frac{{{\phi ^2}\left( r \right)}}{{{f^2}\left( r \right)}} + \frac{2}{{{L^2}f\left( r \right)}}} \right)\psi \left( r \right) = 0.
\end{equation}
For the field to be regular at horizon we impose the condition
\begin{equation}\label{eq13}
\phi \left( {{r_0}} \right) = 0.
\end{equation}
Inserting (\ref{eq13}) into (\ref{eq12}) and expand near $r \to {r_0}$ we arrive at the condition at horizon for scalar field
\begin{equation}\label{eq14}
\frac{{\psi '\left( {{r_0}} \right)}}{{\psi \left( {{r_0}} \right)}} =  - \frac{1}{{2\pi {L^2}T}}.
\end{equation}
At the AdS boundary, the fields $\psi \left( r \right)$ and $\phi \left( r \right)$ behave like  
\begin{equation}\label{eq15}
\psi \left( r \rightarrow\infty \right) = \frac{{{\psi _1}}}{r} + \frac{{{\psi _2}}}{{{r^2}}} + ...,
\end{equation}
\begin{equation}\label{eq16}
\phi \left( r \rightarrow \infty \right) = \mu  - \frac{\rho }{r} + ...,
\end{equation}
where $\mu $ is the chemical potential and the density  $\rho $ associated with the expectation value of charge density, $\rho  = \left\langle {{J^0}} \right\rangle $, with the source term in the boundary action of the form
\begin{equation*}
{S_{bdy}} = {S_{bdy}} + \mu \int {{d^4}} x{J^0}\left( x \right).
\end{equation*}
The holographic duality indicates that there are two possibilities for identifying the sources and the condensates of the dual field theory\\
\begin{itemize}
    \item 
${\psi _1}$ is the source which vanishes at infinity
\begin{equation}\label{eq17}
{\psi _1} = 0,
\end{equation}
and ${\psi _2}$ is condensate ${\psi _2} \approx \left\langle {{O_2}} \right\rangle $.\\

\item ${\psi _2}$ is the source which vanishes at infinity

\begin{equation}\label{eq18}
{\psi _2} = 0,
\end{equation}
then ${\psi _1}$ is condensate ${\psi _1} \approx \left\langle {{O_1}} \right\rangle $.
\end{itemize}

\subsection{Free energy }

In Ref. \cite{phat2021holographic} we found the expressions of free energy corresponding to different quantizations:

-	For ${O_1}$ quantisation the normalised free energy of the boundary theory reads
\begin{equation}\label{eq19}
{\Omega _1} = {V_2}\left[ { - \frac{{\mu \rho }}{2} - \frac{{{O_1}{O_2}}}{{{L^2}}} + \int\limits_0^z {dz\frac{{{\psi ^2}\left( z \right){\phi ^2}\left( z \right)}}{{{z^4}f\left( z \right)}}} } \right],
\end{equation}
where $ z = 1/r $ and ${V_2}$ is the volume of the two – sphere with radius $1/\sqrt k  = 1$.

-	For ${O_2}$ quantization we normalized free energy of the boundary theory takes the analogous form
\begin{equation}\label{eq20}
{\Omega _2} = {V_2}\left[ { - \frac{{\mu \rho }}{2} + \frac{{{O_2}{O_1}}}{{{L^2}}} + \int\limits_0^{{z_0}} {dz\frac{{{\psi ^2}\left( z \right){\phi ^2}\left( z \right)}}{{{z^4}f\left( z \right)}}} } \right],
\end{equation}
which  is the Legendre transform of ${\Omega _1}$.

From (\ref{eq19}) and (\ref{eq20}) we get
\begin{equation*}
\frac{1}{{{V_2}}}\frac{{\partial {\Omega _1}}}{{\partial {O_1}}} =  - {O_2},
\quad 
\frac{1}{{{V_2}}}\frac{{\partial {\Omega _2}}}{{\partial {O_2}}} = {O_1},
\end{equation*}
which tell  that the local extremum of  ${\Omega _1}\left( {{\Omega _2}} \right)$ locates at vanishing ${O_2}\left( {{O_1}} \right)$. 
This  fact is comparable with the assumption  that only one of ${\psi _1}$ and ${\psi _2}$ is non – vanishing for physical solutions.

The free energy corresponding to non – condensed state is given by
\begin{equation}\label{eq21}
\frac{{{\Omega _0}}}{{{V_2}}} = \frac{{{\mu ^2}{z_0}}}{2}.
\end{equation}
Finally the free energy difference reads
\begin{equation}\label{eq22}
\frac{{\Delta {\Omega _i}}}{{{V_2}}} = \frac{{{\Omega _i}}}{{{V_2}}} - \frac{{{\mu ^2}{z_0}}}{2},
\end{equation}
which is what we expect.

\subsection{Conductivity}

Proceeding to the conductivity of the superconductor in the dual CFT as function of frequency, 
we have to solve the equation for the fluctuations of the vector potential ${A_x}\left( {r,t} \right)$in the bulk.
Suppose this potential takes the form
\begin{equation*}
{A_x}\left( {r,t} \right) = {A_x}\left( r \right)\exp \left( { - i\omega t} \right),
\end{equation*}
which leads to the equation of ${A_x}\left( r \right)$ in our set up
\begin{equation}\label{eq23}
{A''_x}\left( r \right) + \frac{{f'\left( r \right)}}{{f\left( r \right)}}{A'_x}\left( r \right) + \left( {\frac{{{\omega ^2}}}{{{f^2}{{\left( r \right)}^{}}}} - \frac{{2{\psi ^2}\left( r \right)}}{{f\left( r \right)}}} \right){A_x}\left( r \right) = 0,
\end{equation}
which, in the new variable $ z = 1/r $,  reads
\begin{equation}\label{eq24}
{A''_x}\left( z \right) + \left( {\frac{2}{z} + \frac{{f'\left( z \right)}}{{f\left( z \right)}}} \right){A'_x}\left( z \right) + \left( {\frac{{{\omega ^2}}}{{{z^4}{f^2}\left( z \right)}} - \frac{{2{\psi ^2}\left( z \right)}}{{{z^4}f\left( z \right)}}} \right){A_x}\left( z \right) = 0,
\end{equation}
where $ \psi $ was solved in subsection $ A $. 
The solution to Eq. (\ref{eq24}) needs to have two boundary conditions. 
The first one is the ongoing condition at horizon
\begin{equation}\label{eq25}
{A_x}\left( z \right) \approx {\left( {z - z_h} \right)^{ - i\omega /4\pi T}} + ...
\end{equation}
with $z_h=1/r_0$.
The second boundary condition is imposed at large $ r $ or, alternatively, at $z=\frac{1}{r} \to 0$
\begin{equation}
{A_x}\left( z \right) = {a_x} + z{b_x} + ...
\label{eq26}
\end{equation}
%
The two boundary conditions, Eqs. (\ref{eq25}) and (\ref{eq26}), allow us to solve the system of two differential equations 
(\ref{eq23}) and (\ref{eq24}).
According to the AdS/CFT duality dictionary  ${b_x}$ determines the boundary current  ${b_x} = {j_x}$. 
The conductivity $\sigma \left( \omega  \right)$ is then derived from the Ohm law 
\begin{equation}\label{eq27}
\sigma \left( \omega  \right) = \frac{{{j_x}}}{{{E_x}}} =  - \frac{{i{b_x}}}{{\omega {a_x}}}.
\end{equation}
The numerical computation will provide the behaviors of the frequency dependent conductivity corresponding respectively to the phase diagrams in the next Section.

\section{Triplet of holographic phase transitions}\label{3}

The problem we solve in this section is to consider 
how the holographic phase transitions change in the process of phase transition 
from small to large BHs in the boundary. 
At first we determine numerically the manifestations of ground state and several excited states which depend on  the value of the chemical $ \mu $.

\subsection{Small BH phase \texorpdfstring{$L = 1$}{L1}  (\texorpdfstring{$P > P_c$}{P gt Pc}) }

\begin{figure}[h] 
	\centering 
	\subfigure[]{ 
		\includegraphics[width= 5.6cm]{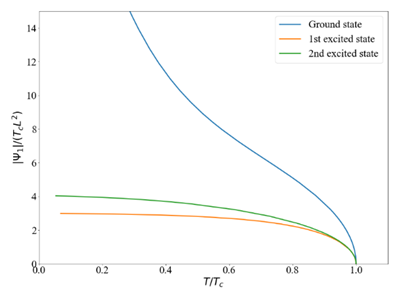}
		\label{f1a}} 
	\subfigure[]{ 
		\includegraphics[width=5.6cm]{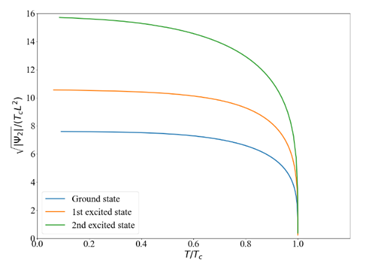}
		\label{f1b}}              
	\caption{The condensations of  ${O_1}$ at $\mu  = 8$ (a) and ${O_2}$ at $\mu  = 12$ (b).} \label{f1}
\end{figure}

Let us begin with the small BH case, which corresponds to the "liquid" side of the vdW transition. 
At $\mu  = 8$, there exhibit a triplet of condensates consisting of ground state (GS) together with first excited state (ES1) 
and second excited state (ES2) for ${O_1}$. 
Their condensations are plotted in Fig. \ref{f1a}, 
where the blue, yellow and green lines indicate GS, ES1 and ES2, respectively. 
The onset of these phase transitions takes place at ${T_c} = 1.6993 {\mu ^{1/2}}$, 0.2952 ${\mu ^{1/2}}$, and 0.0806 ${\mu ^{1/2}}$.

\begin{figure}[h] 
	\centering 
	\subfigure[]{ 
		\includegraphics[width= 5.6cm]{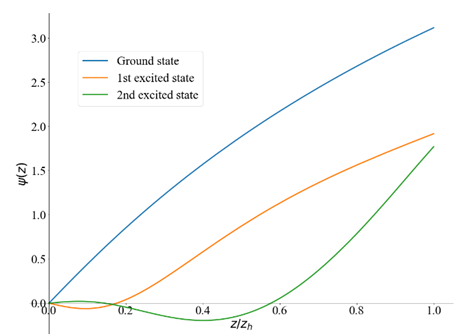}
		\label{f2a}} 
	\subfigure[]{ 
		\includegraphics[width=5.6cm]{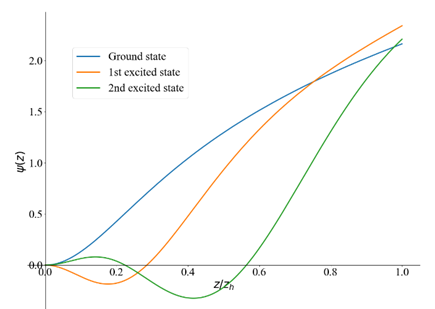}
		\label{f2b}}              
	\caption{The evolutions of  the triplet of condensations versus $ z $.} \label{f2}
\end{figure}

Analogously, at $\mu  = 12$ there emerge also the triplet GS, ES1 and ES2 for ${O_2}$,
and the corresponding phase transitions  are  shown in  Fig.\ref{f1b}.
Their critical temperatures are  ${T_c} =$ 0.7164 ${\mu ^{1/2}}$, 
0.3195 ${\mu ^{1/2}}$, and 0.1532 ${\mu ^{1/2}}$.
The  basic distinction between GS and ES1, ES2 is characterized by their evolutions versus $z$ at corresponding values of chemical potentials.
In Figs. \ref{f2a} and \ref{f2b}, 
these $z$ evolutions of all the foregoing condensations are plotted.

\begin{figure}[h] 
	\centering 
	\subfigure[]{ 
		\includegraphics[width= 5.6cm]{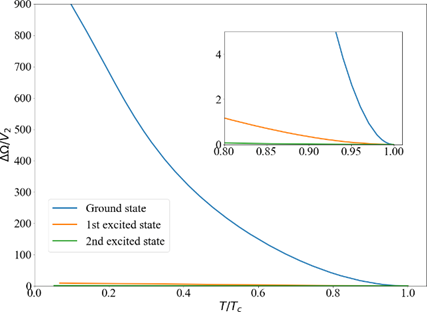}
		\label{f3a}} 
	\subfigure[]{ 
		\includegraphics[width=5.6cm]{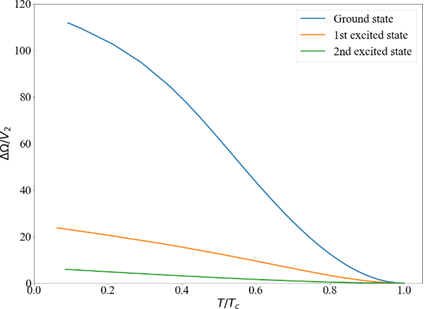}
		\label{f3b}}              
	\caption{The free energy difference for ${O_1}$ ( ${O_2}$ ) is plotted in Fig.\ref{f3a} ( Fig.\ref{f3b} ).} \label{f3}
\end{figure}

In order to determine the order of the foregoing phase transitions, let us calculate numerically the free energy difference, Eq. (\ref{eq22}), 
between the condensed and non-condensed phases for ${O_1}$  and ${O_2}$, respectively. 
They are presented in Fig.\ref{f3a} and \ref{f3b} which prove that 
all phase transitions are of second order.
After fitting the curves in Figs. \ref{f1} near $ T = {T_c} $ we obtain approximately the expressions for triplet of  ${O_1}$ condensates
\begin{equation*}
\avg{O_{1}} = \sqrt 2 {\psi _{1}} \approx {a_{i}}T_c^i{\left( {1 - T/T_c^i} \right)^{1/2}},
\end{equation*}
with ${a_{i}}$ and $T_c^i$  given in the table:
\begin{center}
\begin{tabular}{|m{10em}|m{3cm}|m{2cm}|} 
	\hline
	State  & ${T_c^i}$ & ${a_{i}}$ \\ 
	\hline
	i = 1  GS & 0.2119  ${\mu ^{1/2}}$  & 14.4319\\ 
	\hline
	i = 2  ES1 & 0.04692  ${\mu ^{1/2}}$ & 2.19965\\ 
	\hline
	i = 3 ES2 & 0.04129   ${\mu ^{1/2}}$ & 0.7716 \\ 
	\hline
\end{tabular}
\end{center}

\begin{figure}[h] 
	\centering 
	\subfigure[]{ 
		\includegraphics[width=6cm]{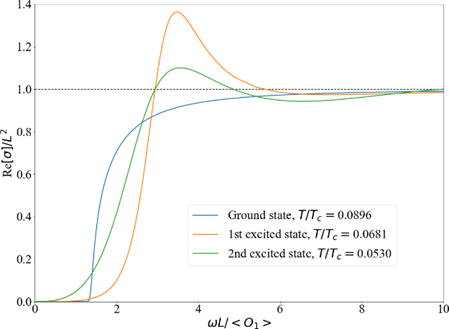}
		\label{f4a}} 
	\subfigure[]{ 
		\includegraphics[width=6cm]{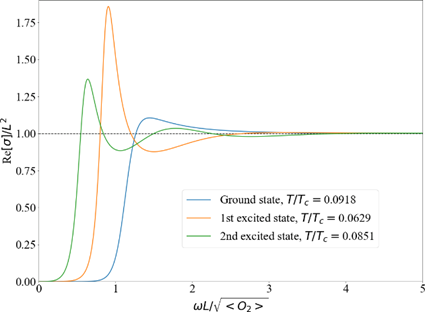}
		\label{f4b}}              
\caption{The real and imaginary parts of optical conductivity 
as function of frequency for ${O_1}$ condensate (a), and ${O_2}$ condensate (b).} 
\label{f4}
\end{figure}

Analogously, for ${O_2}$ we also have the approximate expressions
\begin{equation*}
\sqrt{\avg{O_{2}}}  = \sqrt {2{\psi _{2i}}}  
    = {b_{i}}T_c^i{\left( {1 - T/T_c^i} \right)^{1/2}}.
\end{equation*}
\begin{center}
\begin{tabular}{|m{10em}|m{2cm}|m{2cm}|} 
	\hline
	State  & $T_c^i$ & ${a_{i}}$ \\ 
	\hline
	i = 1  GS & 0.7164 ${\mu ^{1/2}}$ & 17.0127\\ 
	\hline
	i = 2  ES1 & 0.3195 ${\mu ^{1/2}}$  & 9.7429\\ 
	\hline
	i = 3 ES2 & 0.1532 ${\mu ^{1/2}}$  & 6.2474 \\ 
	\hline
\end{tabular}
\end{center}
Finally, we compute numerically the conductivities as functions of the frequency at low temperature for triplets of ${O_1}$ and ${O_2}$. 
They are plotted in Figs. \ref{f4a} and \ref{f4b}. 
The left and right panels correspond respectively to the real and imaginary parts of conductivity.
Fig.\ref{f4a} indicates that GS and ES1 are gapped while ES2 is gapless. 
In Fig.4b we find the poles at $\omega  = 0$ visible in  ${\mathop{\rm Im}\nolimits} \sigma \left( \omega  \right)$. 
Therefore, the real part and imaginary parts  of
$\sigma \left( \omega  \right)$ are related by the Kramers – Kronig relations. 
With regard to the conductivity of  ${O_2}$ we witness the phenomenon similar to that of  ${O_1}$.
The obtained results prove that ES2 in both cases are not the superconducting states,
although their phase transitions are second order.


\subsection{At critical point \texorpdfstring{$L = 6$}{L6}  (\texorpdfstring{$P = P_c$}{P eq Pc}) }

\begin{figure}[h] 
	\centering 
	\subfigure[]{ 
		\includegraphics[width= 6cm]{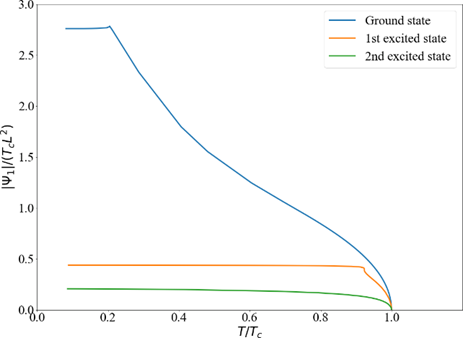}
\label{f5a}} 
	\subfigure[]{ 
		\includegraphics[width=6cm]{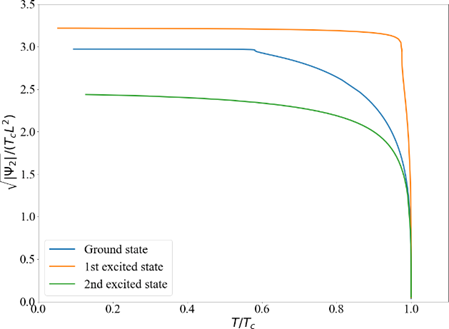}
\label{f5b}}              
\caption{The condensed solution for ${O_1}$ condensate at $\mu  = 1$ (a),
and for ${O_2}$ condensate at $\mu  = 1.2$ (b).}
\label{f5}
\end{figure}

\begin{figure}[h] 
	\centering 
	\subfigure[]{ 
		\includegraphics[width= 6cm]{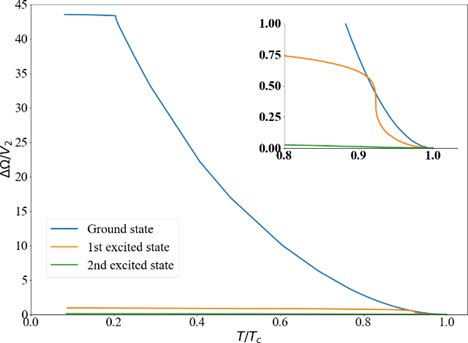}
		\label{f6a}} 
	\subfigure[]{ 
		\includegraphics[width= 6cm]{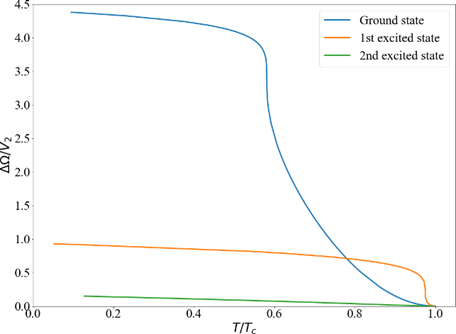}
		\label{f6b}}              
	\caption{The free energy differences are not analytic at corresponding critical temperatures of ES1 for ${O_1}$ ( Fig. \ref{f6a}) and ${O_2}$ ( Fig. \ref{f6b}).} \label{f6}
\end{figure}

The temperature dependence of  GS, ES1 and ES2 condensates are shown in  Figs. \ref{f5a} for  ${O_1}$ at $\mu  = 1$,
and in Fig. \ref{f5b} for ${O_2}$ at $\mu  = 1.2$. 
Their dependence on $ z $ are ignored from now on since they are totally similar to those  presented in Fig. \ref{f2}. 
We next proceed to the free energy difference of above condensates,
which are displayed in Fig. \ref{f6}. 
It is clear that GS and ES2 are second order while ES1 become first order because the free energies differences corresponding to ES1 of  ${O_1}$ and ${O_2}$ are not analytical at critical temperatures. 
Remember that L = 6 corresponds exactly to the critical pressures ${P_c}$ of the small to large BHs phase transition. 
After fitting the curves in Figs. \ref{f5a} and \ref{f5b} near the corresponding temperatures, we get
\begin{equation*}
\avg{O_{1i}} \approx {a_{i}}T_c^i{\left( {1 - T/T_c^i} \right)^{1/2}}
\end{equation*}
with ${a_{i}}$ and $T_c^i$  given in the table
\begin{center}
\begin{tabular}{|m{10em}|m{3cm}|m{2cm}|} 
	\hline
	State  & ${T_c}$ & ${a_{ik}}$ \\ 
	\hline
	GS & 0.2119 ${\mu ^{1/2}}$ & 14.4319\\ 
	\hline
	ES2 & 0.04128 ${\mu ^{1/2}}$  & 0.7716 \\ 
	\hline
\end{tabular}
\end{center}
and for ${O_2}$ we find analogous expressions
\begin{equation*}
{O_{2i}} \approx {a_{i}}T_c^i{\left( {1 - T/T_c^i} \right)^{1/2}}.
\end{equation*}
where ${a_{i}}$ and $T_c^i$  are given in the table
\begin{center}
\begin{tabular}{|m{10em}|m{3cm}|m{2cm}|} 
	\hline
	State  & ${T_c}$ & ${a_{ik}}$ \\ 
	\hline
	GS & 0.0745 ${\mu ^{1/2}}$ & 30.5147\\ 
	\hline
	ES2 & 0.0410 ${\mu ^{1/2}}$  & 14.0848 \\ 
	\hline
\end{tabular}
\end{center}

\begin{figure}[h] 
	\centering 
	\subfigure[]{ 
		\includegraphics[width= 6cm]{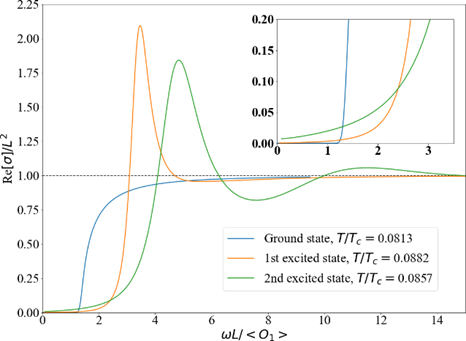}
		\label{f7a}} 
	\subfigure[]{ 
		\includegraphics[width= 6cm]{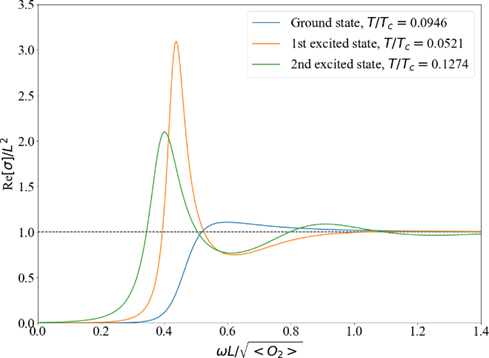}
		\label{f7b}}              
\caption{The real and imaginary parts of conductivities 
for ${O_1}$ condensate (a)  and ${O_2}$ condensate (b).}
\label{f7}
\end{figure}

Let us finally calculate numerically the real and imaginary parts of conductivities for ${O_1}$  and ${O_2}$. 
They are plotted respectively in Fig.\ref{f7a} and  Fig.\ref{f7b}. What we see in both cases is that GS and ES1 are gapped and ES2 is gapless.
In addition, the poles at $\omega  = 0$ are visible in their imaginary parts.
Combining  Fig. \ref{f6a} and \ref{f6b} leads to the conclusion that GS is solely the superconducting state in both cases.
\begin{figure}[h] 
\centering 
\subfigure[]{ 
	\includegraphics[width= 6cm]{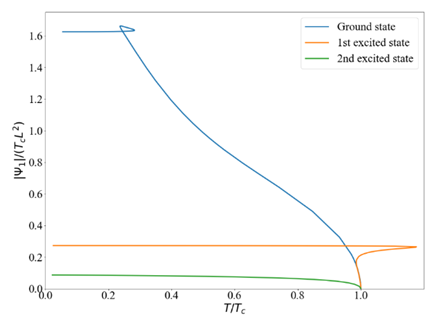}
	\label{f8a}} 
\subfigure[]{ 
	\includegraphics[width=6cm]{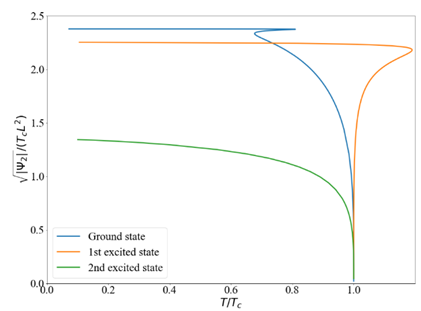}
	\label{f8b}}              
\caption{The temperature dependence of operator expectation values for the ${O_1}$ (${O_2}$) condensate at large $L=9$ are plotted on the left (right) subplot.} 
\label{f8}
\end{figure}


\subsection{Large BH phase \texorpdfstring{$L = 9$}{L9}  (\texorpdfstring{$P < P_c$}{P lt Pc}) }

The $T$ evolutions of GS , ES 1 and ES 2  condensates  are plotted in Fig. \ref{f8a} for ${O_1}$ at $ \mu  = 0.6 $ and Fig. \ref{f8b} for ${O_2}$  at $\mu  = 0.7$.

\begin{figure}[h] 
	\centering 
	\subfigure[]{ 
		\includegraphics[width= 5.2cm]{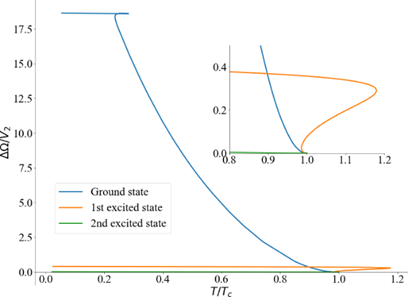}
		\label{f9a}} 
	\subfigure[]{ 
		\includegraphics[width=5.2cm]{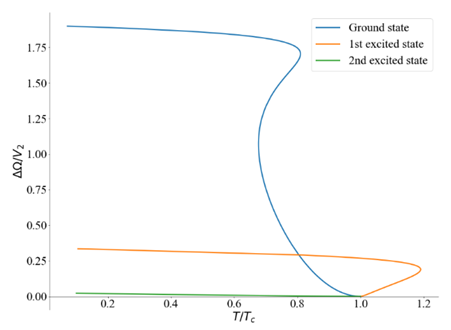}
		\label{f9b}}              
	\caption{The free energy differences for triplet ${O_1}$  (triplet ${O_2}$) are plotted in Fig. \ref{f9a}( Fig. \ref{f9b}).} \label{f9}
\end{figure}

\begin{figure}[h] 
	\centering 
	\subfigure[]{ 
		\includegraphics[width= 6.0cm]{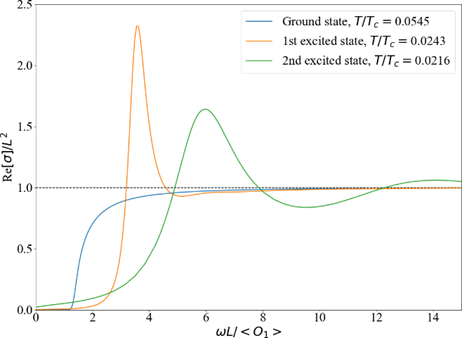}
		\label{f10a}} 
	\subfigure[]{ 
		\includegraphics[width=6.0cm]{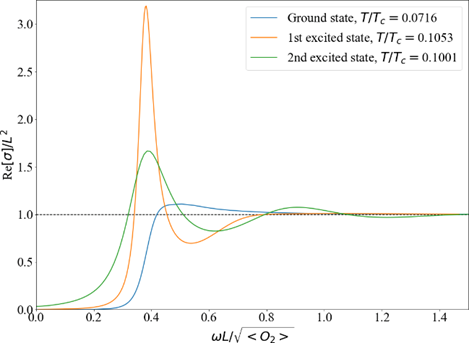}
		\label{f10b}}              
\caption{The real part of the conductivity ${\mathop{\rm Re}\nolimits}~ \sigma \left( \omega  \right)$ 
plotted as functions of $\omega /T$ for ${O_1}$ condensate (Fig.\ref{f10a})
and for ${O_2}$ condensate (Fig. \ref{f10b}).}
\label{f10}
\end{figure}

The free energy differences for condensates ${O_1}$  and ${O_2}$ are plotted respectively in Fig. \ref{f9a} and Fig. \ref{f9b} which also confirm that GS and ES2 are second order while ES1 is first order.
Finally, the real and imaginary parts of conductivities of ${O_1}$ and ${O_2}$ are depicted in Fig. \ref{f10a} and Fig. \ref{f10a}, respectively.
They also assert that GS and ES1 are gapped, while ES2 is gapless. 
Thus, similar to the above case, GS is the unique superconducting state.

\section{Conclusion and discussion}\label{4}

The main purpose of this paper is to consider how the triplet of holographic transitions at the boundary changes 
when the BH undergoes a phase transition from small to large BHs in the bulk.
Based on the numerical computations of free energy and conductivity, 
the following effects are observed:
\begin{itemize}

\item 
When the BH in the bulk is large BH, all GS, ES1 and ES 2 are second order phase transitions, and the corresponding states are superconducting.

\item
When the BH in the bulk is large BH the GS is of second order 
while ES1 and ES2 are of the first order.
In this case, GS is solely the superconducting state. 
Thus, the phase transition from small to large BHs in the bulk generated in the bulk a mechanism which makes change the physical properties of holographic phase transitions in the boundary. 
Our setup provides a new type of holographic superconductivity associated directly with topological charge of BH. 
This is the novel feature of our paper.

\end{itemize}

It is worth to remark that there exist the basic difference between our paper and several Refs. \cite{nishioka2010holographic, franco2010holographic, basu2019holographic, hafshejani2019unbalanced}
where the authors introduced different mechanisms which lead to the change of  phase transition order. However, the authors of these papers did not study excited states. 
In \cite{nishioka2010holographic} the holographic phase transition is associated with the Hawking – Page in the bulk because the AdS soliton decays into AdS BH via Hawking – Page transition.

In  Ref. \cite{franco2010holographic} the charged scalar field is forced to condense by another neutral scalar field,
and in \cite{basu2019holographic} the non-linear interaction was employed $F = \sum\limits_{i = 2}^4 {{c_i}} {\left| \psi  \right|^i}$ 
and one found for ${c_4} \ge 1$ the first order phase transition exhibited  and, moreover, the gap becomes narrower as increases from $ 0 $ to $ 1 $.

In \cite{hafshejani2019unbalanced} the unbalanced Stagestruck holographic superconductors was considered. The main result is that depending on the parameters of this model the phase transition also changes from second order to first order and, at the same time, the conductivity gaps are affected strongly.

It is easily recognized that in order to make change the order of holographic phase transitions, the authors of these works had to choose different Lagrangian of scalar fields by hand. Meantime, our paper shows that the phase transition from small to large BHs in the bulk  creates a mechanism which makes change the order of holographic phase transitions in the boundary. Our results provide a deeper view on the relation between the bulk and the boundary in the AdS/CFT duality.

Our numerical results show that the conductivity $\sigma (\omega)$ show gapped behavior for $L=1$. On the other hand, for $L=9$, 
the ground and first excited states show gapped behavior, 
while the second excited states show gapless behavior. 
This correlates with the criticality behavior that we investigated
in a companion paper \cite{nguyen2021asymptotic}.
In that work, it is found that the spectrum of threshold chemical
potentials of condensate states is continuous in the limit $T\rightarrow 0$
for $L=1$. For large $L=9$, this spectrum is discrete for ground
and first excited states, but is continuous for second excited states and higher.

Last but not least we conjecture that some drastic changes also occur in the boundary when the scalar field in the bulk is replaced by other matter fields.


\bibliographystyle{apsrev4-2}


\bibliography{triplet}   

\end{document}